\documentclass[reprint, longbibliography,noeprint]{revtex4-1}
\usepackage{graphicx}% Include figure files
\usepackage{dcolumn}% Align table columns on decimal point
\usepackage{bm}% bold math
\usepackage{amsmath}
\usepackage{amssymb}
\usepackage{cancel}
\usepackage{chemarr}
\usepackage{hyperref}
\usepackage[dvipsnames]{xcolor}
\usepackage{mhchem}

\definecolor{webgreen}{rgb}{0,.5,0}
\definecolor{webbrown}{rgb}{.6,0,0}
\definecolor{grigio}{rgb}{.85,.85,.85} 
\definecolor{RoyalBlue}{rgb}{0.0, 0.14, 0.4}
\definecolor{skyblue1}{rgb}{0.45,0.62,0.81}
\definecolor{skyblue2}{rgb}{0.2,0.39,0.64}
\definecolor{skyblue3}{rgb}{0.13,0.29,0.53}
\definecolor{scarlet1}{rgb}{0.93,0.16,0.16}
\definecolor{scarlet2}{rgb}{0.8,0,0}
\definecolor{scarlet3}{rgb}{0.64,0,0}

\definecolor{g}{gray}{0.50}

\hypersetup{%
    colorlinks=true, linktocpage=true, pdfstartpage=1, pdfstartview=FitV,%
    breaklinks=true, pdfpagemode=UseNone, pageanchor=true, pdfpagemode=UseOutlines,%
    plainpages=false, bookmarksnumbered, bookmarksopen=true, bookmarksopenlevel=1,%
    hypertexnames=true, pdfhighlight=/O,%nesting=true,%frenchlinks,%
    urlcolor=webbrown, linkcolor=RoyalBlue, citecolor=webgreen, 
    pdfsubject={},%
    pdfkeywords={},%
    pdfcreator={pdfLaTeX},%
    pdfproducer={LaTeX REVTeX}%
}

\begin{document}

\preprint{APS/123-QED}

\title{A chemical reaction network implementation of a Maxwell demon}

\author{Massimo Bilancioni}
\email{massimo.bilancioni@uni.lu}
\affiliation{Department of Physics and Materials Science, University of Luxembourg, avenue de la Fa\"{i}encerie, Luxembourg City, 1511 G.D.~Luxembourg}
\author{Massimiliano Esposito}
\email{massimiliano.esposito@uni.lu}
\affiliation{Department of Physics and Materials Science, University of Luxembourg, avenue de la Fa\"{i}encerie, Luxembourg City, 1511 G.D.~Luxembourg}
\author{Nahuel Freitas}
\email{nfreitas@df.uba.ar}
\affiliation{Department of Physics and Materials Science, University of Luxembourg, avenue de la Fa\"{i}encerie, Luxembourg City, 1511 G.D.~Luxembourg}
\affiliation{Universidad de Buenos Aires, Facultad de Ciencias Exactas y Naturales, Departamento de Física. Buenos Aires, Argentina.}

\date{\today}% It is always \today, today,

\begin{abstract}
We study an autonomous model of a Maxwell demon that works by rectifying  thermal fluctuations of chemical reactions. It constitutes the chemical analog of a recently studied electronic demon. 
%notably the fact that it is composed of two modules that can be thought of as chemical inverters.
%Similarities and differences between the two are highlighted.
We characterize its scaling behavior in the macroscopic limit, its performances, and the impact of potential internal delays. We obtain analytical expressions for all quantities of interest, namely, the generated reverse chemical current, the output power, the transduction efficiency, and the correlations between the numbers of molecules.
Due to a bound on the nonequilibrium response of its chemical reaction network, we find that, contrary to the electronic case, there is no way for the Maxwell demon to generate a finite output in the macroscopic limit. 
Finally, we analyze the information thermodynamics of the Maxwell demon from a bipartite perspective. In the limit of a fast demon, the information flow is obtained, its pattern in the state space is discussed, and the behavior of the partial efficiencies related to the measurement and the feedback processes is examined.  

\end{abstract}

\maketitle

%\tableofcontents
\section{Introduction}
\label{sec:intro}

A Maxwell demon (MD) is a thought experiment conceived in 1867 by J. C. Maxwell~\cite{maxwell1871,leff2002} to emphasize the statistical nature of the second law of thermodynamics and to challenge its validity at the microscopic scale. 
In this thought experiment,  a small intelligent being, the demon, seemingly violates the second law by bringing out of equilibrium a gas that was initially in equilibrium without any apparent energetic cost.  It achieves this goal by controlling a microscopic gate and sorting gas particles according to their speeds. This alleged violation of the second law stimulated crucial conceptual advances in the last century due to 
Szilard~\cite{szilard1929}, Brillouin~\cite{brillouin1951}, Landauer~\cite{5392446,landauer1991} and Bennet~\cite{bennett1982} that revolutionized our understanding of thermodynamics by incorporating information into it. What was realized was that there is a fundamental thermodynamic cost associated with the processing of information that allows the demon to function. This cost arises either in the measurement process, in the resetting of the demon's memory or in both steps and it ensures that the second law, despite being statistical, continues to hold even at the microscopic scale. Nowadays, the general MD is understood as an ``information engine" that functions in accordance with the second law~\cite{PhysRevLett.102.250602,PhysRevE.79.041118,PhysRevE.85.021104,Esposito_2012,barato2014}: the demon consumes energy to act on a system as an active feedback-control loop, its operation rectifies the system's fluctuations making it possible to extract work from the latter.
In the last decades, there have been many experimental realizations corroborating this picture. Their physical implementations are diverse: molecular systems~\cite{serreli2007,Alvarez-Pérez2008,Carlone2012-pc,Amano2022}, colloidal particles~\cite{toyabe2010,saha2021}, laser cooling~\cite{Travis_Bannerman_2009,kumar2018}, single-electron circuits ~\cite{koski2014,PhysRevLett.115.260602}, nuclear-spin system~\cite{PhysRevLett.117.240502}, electro-photonic system~\cite{PhysRevLett.116.050401}, superconducting qubit~\cite{cottet2017,Masuyama2018,PhysRevLett.121.030604}, DNA-hairpin~\cite{ribezzi}, and cavity QED setups~\cite{PhysRevResearch.2.032025}.
\\

In the present paper, we thoroughly analyze a  theoretical model of a MD based on chemical reaction networks (CRN): we realized that a recent electronic implementation of a MD~\cite{electronic,electronic_info_flow} can be translated into chemistry conserving the same structure and working principle. In particular, the resulting MD is composed of two modules that can be thought of as ``chemical inverters". This system offers an interesting viewpoint for investigating the analogies and differences between CRN and electronics. 
Moreover, despite not being the first MD implemented with CRN, it differs qualitatively from the previous ones~\cite{serreli2007,Alvarez-Pérez2008,Carlone2012-pc,Amano2022, Flatt2021.12.03.471046, Penocchio_2022}.
In those prior systems, the macroscopic limit is, in the end, equivalent to having many Maxwell demons working in parallel, whereas the same is not true in our case. Our MD's mechanism  shows itself as a rectification of
the thermal fluctuations in the \emph{numbers} of molecules and, since the relative size of these fluctuations goes to zero as the size $V$ of the system increases, the MD effect disappears.
In the paper, we  investigate the scaling behavior of our MD in the limit $V\to\infty$. We perform it with the tools of stochastic thermodynamics~\cite{seifert2012,schmiedl2007,PhysRevX.6.041064} and, in particular, we take advantage of the bipartite formalism~\cite{PhysRevX.4.031015,article} to analyze the Maxwell demon's information thermodynamics.
The resulting analysis provides a lucid understanding of the MD's functioning: each component of its chemical reaction network can be distinctly interpreted in relation to its functionality, and the analytical solvability of the model enables a comprehensive exploration of the MD's performance in different regimes. 
Most results obtained are in line with what was found for the electronic version of the MD~\cite{electronic,electronic_info_flow}, the main difference being that, in chemistry, increasing the input power does not allow one to mantain a finite  MD's output in the macroscopic limit.
 \\

This paper is organized as follows.  Sec.~\ref{sec:CRN} is an introduction to the concepts of chemical reaction networks necessary for understanding the chemical Maxwell demon. In Sec.~\ref{sec:chem_inv}, we present  the chemical inverter and its working principle. In \ref{subsubsec:simplified_chem_inv}, we quantitatively analyze its simplified version employed in the rest of the paper, and, in \ref{sec:comparison}, we highlight the main differences with the electronic inverter. The beginning of Sec.~\ref{sec:chem_dem} is a preliminary overview of the chemical Maxwell demon: we dissect its structure, we sketch its working principle, and we briefly discuss its basic thermodynamics.
In  Sec.~\ref{sec:setup}, we clarify the detailed setup and how the macroscopic limit is performed; we also outline the  procedure followed  to solve the Maxwell demon. In Sec.~\ref{sec:linking correlations}, we show the central role played by the correlations between the number of molecules. Namely, we connect all the other relevant quantities to the covariance between these numbers, which is computed in the subsequent Sec.~\ref{sec:computing correlations}. This is done paying close attention to the accuracy of rate function methods. The covariance's expression is then used to assess the effect of internal delays in the demon and to derive the transduction efficiency  analyzed in Sec.~\ref{sec:efficiency}. 
%We find an upper bound for this efficiency and based on its small value we argue about the plausibility of finding such a mechanism in living cells.\\
Finally, Sec.~\ref{sec:info flow} consists in further resolving the thermodynamics of the Maxwell demon from a bipartite perspective: we explain how the demon-system information flow is incorporated into its thermodynamics and we compute it in the limit of a fast demon. The expression of the information flow allows us to infer the behavior of the partial efficiencies related to the measurement and feedback processes. Lastly, we discuss the information flow's pattern in the state space. 
\section{Chemical reaction networks}
\label{sec:CRN}
\begin{figure*}[htbp!]
\centering
\includegraphics[scale=0.34]{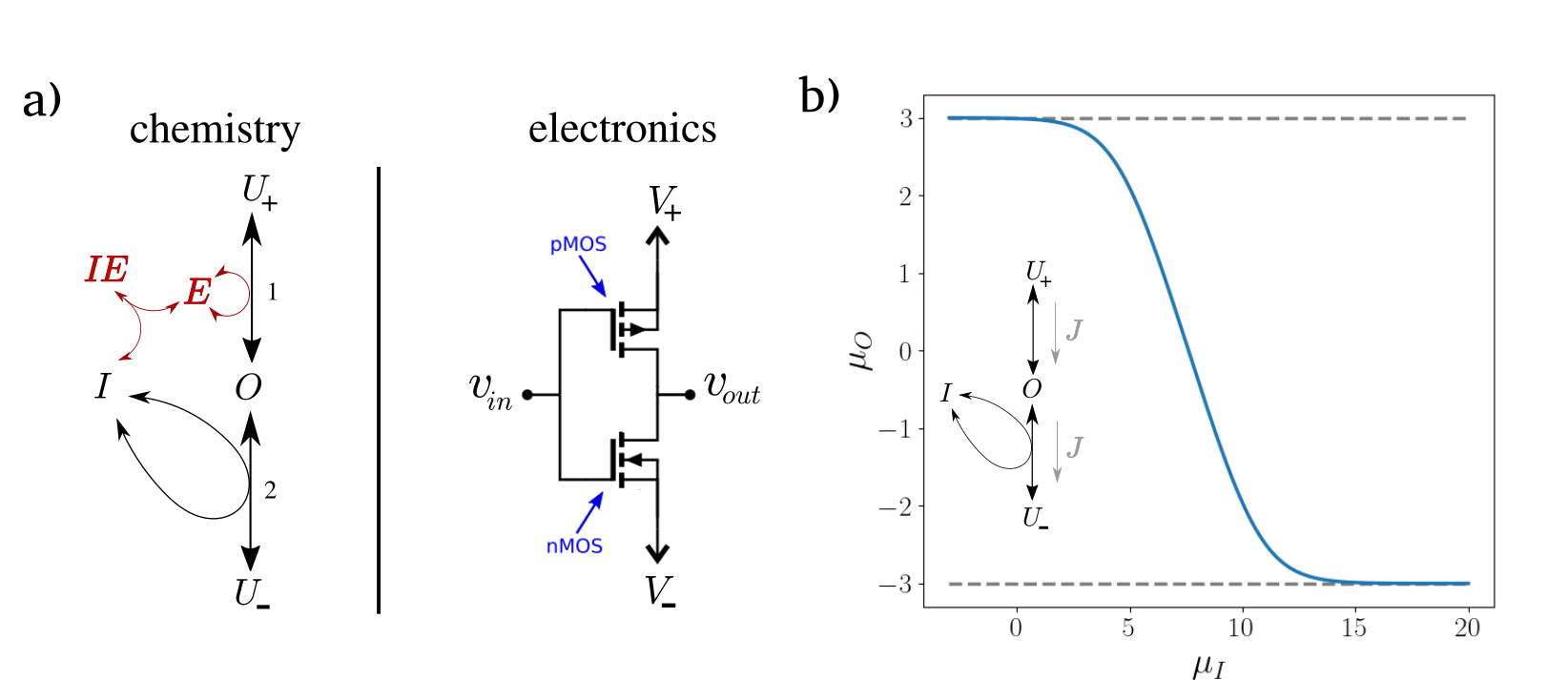}
\caption{Chemical inverter: a) Comparison between the structure of the chemical and the electronic inverter. The black part of the CRN, without the red subnetwork, constitutes the simplified version of the chemical inverter used to build the Maxwell demon. 
b) The input-output relation of the simplified chemical inverter in terms of chemical potentials, Eq.~\eqref{eq:i-o relation}. The dotted lines correspond to the values $\mu_{U_+}$ and $\mu_{U_-}$. The parameters are $\mu_{U_+} = -\mu_{U_-} =3 \,(k_BT) $, $\frac{k_+V}{k_-} =100$.
}
\label{fig:inverter}
\end{figure*}

In this  Section, we briefly introduce the concepts of chemical reaction networks~\cite{schmiedl2007,PhysRevX.6.041064,rao2018} constituting the basis of our  chemical Maxwell demon.  \\
A chemical reaction network (CRN) is defined as a set of species $\{\alpha\}$ and chemical reactions $\{\rho\}$ among these species. We consider as a demonstrative example the network  in Fig.~\ref{fig:inverter} a) (black part) that will be later analyzed in  Sec.~\ref{subsubsec:simplified_chem_inv}:
\begin{equation}
\begin{split}
\ce{ $O$ &<=>[+1][-1] $ U_+$} \\
\ce{$U_-  + I$& <=>[+2][-2] $O+I $ }.
\end{split} 
    \label{eq: example CRN}
\end{equation}
There are four species connected by two chemical reactions; in the second, $I$ represents an enzyme.
Physically, we imagine those species to be dissolved in a much more abundant inert solvent forming an ideal-dilute solution of volume $V$.
We assume that the time-scale in which the solute molecules diffuse is much shorter than that of chemical reactions.
%This assumption is in agreement with considering small sizes $V$. As we will see, this is the limit in which the Maxwell demon can work.
In this way, the solution can be considered homogenous at all times and  the chemical concentrations represent the only out-of-equilibrium degrees of freedom. 
The states $\boldsymbol n$ of the system are the ones corresponding to a given number of molecules for each chemical species,
\begin{equation}
    \boldsymbol n = (n_I,n_{U_+},n_{U_-},n_O).
\end{equation}
The transition among those states occur through chemical reactions  happening stochastically inside the solution. The resulting system's dynamics is a Markovian jump-process.
As we will see, stochasticity is a crucial ingredient for the Maxwell demon to function. 
The stochastic reaction rates $\lambda_\rho$ obey the mass-action law:  they are proportional to the product of the number of  reactant molecules present in the solution,
\begin{equation}
    \begin{split}
     \lambda_{ +1}(\boldsymbol n) = k_{ +1}\, n_{O}   &\quad \lambda_{-1}(\boldsymbol n) = k_{-1}\, n_{U_+}  \\
        \lambda_{+2}(\boldsymbol n) = k_{+2}\frac{n_I n_{U_-}}{V}&\quad \lambda_{-2}(\boldsymbol n) = k_{-2}\frac{  n_{O}n_I}{V},
    \end{split}
       \label{eq: intro CRN rates MA}
\end{equation}
 where $k_{+1},\, k_{ -1},\,
        k_{+2},\, k_{ -2}$ are the rate constants. The volume $V$ shows up to ensure $\lambda_\rho \propto V$. \\
For ideal-dilute solutions, where chemical concentrations are the only out-of-equilibrium degrees of freedom, the nonequilibrium chemical potential can be written as a function of the concentration in the form~\cite{PhysRevX.6.041064,alberty2003}
\begin{equation}
    \mu_\alpha = \mu_\alpha^0(T) + k_BT \log \frac{\langle n_\alpha \rangle}{V}.
    \label{eq: chem pot form}
\end{equation}
The higher the concentration $\frac{\langle n_\alpha \rangle}{V}$, the higher the chemical potential of $\alpha$. $\mu_\alpha^0(T)$ represents the standard chemical potential which depends on the solvent's nature and on the temperature $T$.\\
For an isolated CRN, the Gibbs free energy
\begin{equation}
    G = \sum_\alpha \mu_\alpha n_\alpha
\end{equation}
decreases over time until the equilibrium value is reached~\cite{PhysRevX.6.041064}. This means that chemical reactions spontaneously proceed  in the direction that reduces the chemical potential difference between reactants and products. When equilibrium is reached,
for every reaction, the sum of the chemical potentials of the reactants must be equal to the sum of  the chemical potentials of the products, which means, from Eq.~\eqref{eq: example CRN},
\begin{equation}
    \begin{split}
       \mu_{O}^{eq} = \mu_{U_+}^{eq} \quad\quad
         \mu_{U_-}^{eq}=\mu_{O}^{eq}.
    \end{split}
\end{equation}
If one combines this requirement with the fact that the equilibrium dynamics is detailed-balance and thus there are no net chemical currents,
\begin{equation}
    \begin{split}
         J_{O \to U_+} = \langle \lambda_{ +1}(\boldsymbol n) - \lambda_{ -1}(\boldsymbol n)\rangle = 0\\
         J_{ U_-\to O}  = \langle \lambda_{ +2}(\boldsymbol n) - \lambda_{ -2}(\boldsymbol n)\rangle= 0,
    \end{split}
\end{equation}
one can derive the local detailed-balance conditions~\cite{rao2018}
\begin{equation}
\begin{split}
    \log \left(\frac{k_{+1}}{k_{ -1}}\right) =  \mu_{O}^0- \mu_{U_+}^0\\
     \log \left(\frac{k_{+2}}{k_{ -2}}\right) = \mu_{U_-}^0-\mu_{O}^0.
\end{split}
\label{eq:ldb}
\end{equation}
This condition
guarantees the model's thermodynamic consistency and it remains valid also for open CRNs.\\

Now, we move on to discuss open CRNs that can exchange chemical species with the environment. The CRN analyzed in this paper falls into this category and the reason for considering this kind of CRNs is that it simplifies the treatment: it allows one to remove  some species from the dynamics by externally fixing their concentrations. We can imagine  the system to be connected with chemical reservoirs, known as chemostats. Each chemostat can promptly exchange only one  species with the system. Its size is assumed to be much bigger than the size of the system so that any perturbation induced by the evolution of the latter is effectively negligible. 
By replenishing or taking away molecules, the net effect of each chemostat is to held the concentration (or equivalently the chemical potential) of the exchanged species constant  in the system and equal to the chemostat's value.\\
For example, in the CRN of Eq.~\eqref{eq: example CRN}, we can imagine to externally fix the concentration of the species $U_-,\,U_+$ by connecting the solution to two chemostats with chemical potentials $\mu_{U_-},\, \mu_{U_+}$. In the case $\mu_{U_+}> \mu_{U_-}$, the system  reaches a nonequilibrium steady state with a nonzero positive current $J_{U_+\to U_-}$ flowing between the two chemostats. In the steady state, the free energy of the system is constant over time. However, free energy is being extracted at a rate $\dot{\mathcal{F}}_{U_+} = -J_{U_+\to U_-}\,\mu_{U_+}$ in the $U_+$ chemostat and free energy is being released at a rate $\dot{\mathcal{F}}_{U_-} = J_{U_+\to U_-}\,\mu_{U_-}$ in the $U_-$ chemostat. The overall free energy consumed   in stationary conditions is equal to the dissipation or entropy production rate~\cite{seifert2012,rao2018}
\begin{equation}
    T\dot \sigma  =-\dot{\mathcal{F}}=-(\dot{\mathcal{F}}_{U_+}+\dot{\mathcal{F}}_{U_-}) =  J_{U_+\to U_-}\Delta\mu
    \label{eq: dissipation chemostats}
\end{equation}
with $\Delta\mu = \mu_{U_+} -\mu_{U_-}$.
\\

In the following, all chemical potentials and free energies will be measured in units of $k_BT$.
	
\section{Chemical inverter}
\label{sec:chem_inv}

As mentioned in the introduction, the Maxwell demon is composed of two equivalent modules linked together. These modules can be thought of as chemical inverters. A good starting point for understanding the Maxwell demon is to follow a modular approach~\cite{avanzini}: we first introduce and characterize the properties of a chemical inverter alone. \\
The CRN corresponding to the chemical inverter is shown in Fig.~\ref{fig:inverter} a).

%It is composed by six chemical species connected by three reactions
%\begin{equation}
 %   \begin{matrix}
  %      I +E\xrightleftharpoons{} IE \\
      %U_+ +E\xrightleftharpoons{} O+E  \\
      %O+I\xrightleftharpoons{}  U_-  + I
%     \end{matrix}
 %    \label{eq: example CRN}
 %\end{equation}
Its structure and working principle take inspiration from the CMOS electronic inverter~\cite{electronic} shown in the same figure. In the electronic case, the input potential $v_{\text{in}}$ controls the output potential $v_{\text{out}}$. It does so through p-MOS and n-MOS transistors that act on the conductivity of the two channels connecting $v_{\text{out}}$ to the terminal potentials $V_+$ and $V_-$, with $V_+>V_-$. The name ``inverter" stems from the fact that when $v_{\text{in}}$ is high, $v_{\text{out}}$ is low and vice versa. Indeed, if $v_{\text{in}}$ is high, what happens is that the p-MOS transistor blocks the upper channel, while the n-MOS transistor makes the lower channel very conductive, resulting in $v_{\text{out}}\simeq V_-$. If $v_{\text{in}}$ is low, the opposite happens, resulting in $v_{\text{out}}\simeq V_+$.\\
In the chemical case,
the input species is $I$, the output species is $O$, there are two internal species $E,EI$ and two terminal species $U_+, U_-$ that can interconvert into $O$ through the upper and lower chemical reactions. In this context, the reaction rates and chemical potentials play, respectively, the roles of the conductivity and electric potentials. The terminal species $U_+$ and $U_-$  are considered to be chemostatted (Sec.~\ref{sec:CRN}) to chemical potentials  $\mu_{U_+}$ and $\mu_{U_-}$, with $ \mu_{U_+}>\mu_{U_-}$. The input potential $\mu_I $ controls the output potential $\mu_O $ in an inverter-like manner  by acting on the speed of the lower ($O\xrightleftharpoons{}  U_- $) and upper ($ U_+  \xrightleftharpoons{} O $) chemical reactions. In the first case, this influence is made possible via the enzymatic reaction 
 \begin{equation}
     O+I\xrightleftharpoons{}  U_-  + I.
 \end{equation}
The higher $\mu_I$ and therefore the concentration $\langle n_I\rangle/V$ (Eq.~\eqref{eq: chem pot form}), the higher the rate at which this reaction takes place. This reproduces in chemistry the qualitative behavior of the n-MOS transistor: a higher $v_{\text{in}}$ results in a more conductive lower channel. In the second case, $\mu_I$ affects the upper reaction through the subnetwork 
 \begin{equation}
    \begin{split}
    I + E  &\xrightleftharpoons{}  IE\\
      U_++ E &\xrightleftharpoons{} O +E.
     \end{split}
 \end{equation}
 Whenever  $\mu_I$ is high, the enzyme $E$ is converted into the inactive species $IE$, thus, slowing down the interconversion $ U_+  \xrightleftharpoons{} O $. This behavior represents the chemical analogue of the p-MOS transistor: a higher $v_{\text{in}}$ results in a less conductive upper channel. \\
 Putting everything together, if $\mu_I$  is high, the lower reaction is much faster than the upper one. This translates into $O$ almost being in equilibrium with the lower species $U_-$, which means $\mu_O \simeq \mu_{U_-}$. If, on the other hand, $\mu_I$  is low, the reverse happens and we have $\mu_O \simeq \mu_{U_+}$. We conclude that the CRN of Fig.~\ref{fig:inverter} a) exhibits the qualitative inverter-like behavior.\\
We stress that both  the electrical and chemical inverters must dissipate free energy in order to function. As a matter of fact, they need a nonzero potential difference applied to their terminals to be able to respond to an input signal. This potential difference translates, at steady state, into a nonzero electrical/chemical current flowing
 from the higher to the lower electrical/chemical potential leading to a certain dissipation, which is  given by Eq.~\eqref{eq: dissipation chemostats} in the chemical case.
 \\
A final consideration is that  the CRN of Fig.~\ref{fig:inverter} a) still behaves as an inverter even without the ``p-MOS subnetwork" (the red portion) \footnote{ An electronic analog of this simplified version can be identified in the p-MOS inverter with load resistor~\cite{nair2002}.}. This reduced CRN is simpler  since there are no internal species and less symmetric because the input potential $\mu_I$ now only affects the speed of the lower reaction. Nevertheless, this simplified inverter is still sufficient for our goal of building a Maxwell demon and it greatly simplifies its analysis. Therefore, in the rest of the paper, we will consider this version.

\subsection{Analysis of the simplified chemical inverter}
\label{subsubsec:simplified_chem_inv}
The CRN corresponding to the simplified chemical inverter is the one already presented in Eq.~\eqref{eq: example CRN}: 
\begin{equation}
    \begin{split}
      U_+ &\xrightleftharpoons{} O \\ O+I &\xrightleftharpoons{}  U_-  + I.
     \end{split}
     \label{eq: CRN simplified inverter}
 \end{equation}
 In this Section, we quantitatively characterize its behavior. In particular,  we find its steady state input-output relation. To this aim, we imagine that the input species $I$ and the terminal species $U_-,U_+$ are chemostatted respectively to chemical potentials $\mu_I$, $\mu_{U_-}$, $\mu_{U_+}$ (Sec. \ref{sec:CRN}). 
  For convenience of notation, we filter out standard chemical potentials by setting  $\mu_O^0 = \mu_I^0 =0 $. This is equivalent to  measure $\mu_I$ relative to $\mu_I^0 $  and $\mu_O$, $\mu_{U_-}$, $\mu_{U_+}$ relative to $\mu_O^0 $. In this way,
  \begin{equation}
    \begin{split}
    \mu_I= \log \frac{\langle n_I\rangle}{V} \quad  \quad 
        \mu_{U_-}= (\mu_{U_-}^0 -\mu_{O}^0)  + \log \frac{\langle n_{U_-}\rangle}{V} \\
         \mu_O = \log \frac{\langle n_O\rangle}{V}\quad  \quad
        \mu_{U_+}= (\mu_{U_+}^0 -\mu_{O}^0)  + \log \frac{\langle n_{U_+}\rangle}{V}.
    \end{split}
    \label{eq: chem pot without standard}
\end{equation}
We assign the rate constants so that they are in compliance with the local detailed-balance conditions~\eqref{eq:ldb}:
\begin{equation}
\begin{split}
     k_{ +1} = k_+\quad k_{-1} = \exp(\mu_{U_+}^0- \mu_{O}^0)\, k_+ \\
 k_{ -2} = k_-\quad k_{+2} = \exp(\mu_{U_-}^0- \mu_{O}^0)\, k_-.
\end{split}
\label{eq: rate constants assigned}
\end{equation}
By substituting expressions~\eqref{eq: rate constants assigned} and \eqref{eq: chem pot without standard} into Eq.~\eqref{eq: intro CRN rates MA} we get the stochastic rates:
\begin{equation}
    \ce{ $U_-$ <=>[$ k_- V\exp(\mu_{U_-} +\mu_I )$][$k_-n_O \exp(\mu_I)$] $O$ <=>[$k_+ n_O$][$ k_+ V\exp(\mu_{U_+})$] $U_+  $ }.
    \label{eq:rate choice simplified inverter}
    \end{equation}
    The input-output relation is then obtained by 
the steady state condition, which requires the two average chemical currents  to be equal 
\begin{equation}
\begin{split}
J = J_{ U_-\to O} &= J_{O \to U_+}\\
k_-e^{\mu_I} \left( Ve^{\mu_{U_-}} -\langle n_O\rangle\right) &= k_+\left(\langle n_O\rangle -V e^{\mu_{U_+}}\right).
\end{split}
\end{equation}
This leads to
    \begin{equation}
    \mu_O= \log \frac{\langle n_O\rangle}{V} =\log \left[\frac{k_{+}e^{\mu_{U_+} } +k_{-} e^{\mu_I +\mu_{U_-} } } {k_{+}+k_{-} e^{\mu_{I}}}\right].
    \label{eq:i-o relation}
\end{equation}
Keeping in mind that $e^{\mu_I}$ is the concentration, we can recognize inside the parenthesis the Hill function~\cite{segel1989} with Hill coefficient equal to 1. In Fig.~\ref{fig:inverter} b), we plot this input-output relation. The behavior is the one expected for an  inverter: for high $\mu_I$, $\mu_O\simeq \mu_{{U_-}}$, while for low $\mu_I$,  $\mu_O\simeq \mu_{{U_+}}$. In the middle, there is a window of $\mu_I$ over which the inverter transitions between these two limiting cases. At steady state, to function, the inverter dissipates free energy from its chemostats at a rate given by Eq.~\eqref{eq: dissipation chemostats}
\begin{equation}
    T \dot \sigma = J\Delta\mu = V\frac{k_+k_-e^{\mu_I} ( e^{\mu_{U_+}} - e^{\mu_{U_-}})}{k_+ +k_-e^{\mu_I}} \Delta \mu.
\end{equation}
The dissipation scales $\propto V$, the system size.\\
Finally, we mention in advance two more properties that  will be used in the analysis of the Maxwell demon. They come from the fact that the stochastic process depicted in Eq.~\eqref{eq:rate choice simplified inverter} is a 1D linear jump-process~\cite{gardiner2009}. Firstly, the relaxation rate for the mean $\langle n_O\rangle$  is given by
\begin{equation}
    k_{\text{relax}} = k_+ +k_-e^{\mu_I}.
    \label{eq:relax_rate}
\end{equation}
Secondly, 
the steady state distribution $P_{\rm ss}(n_O)$ is 
Poissonian~\cite{gardiner2009} implying for the fluctuations  
\begin{equation}
    \langle\Delta n_O^2\rangle = \langle n_O\rangle.
\end{equation}
\subsection{Chemical \emph{vs.} electronic inverter}
\label{sec:comparison}
Before moving to the Maxwell demon, we go a little deeper in the comparison between the two kinds of inverter. 
From a physical point of view, we already mentioned that
the analogues of electric potentials and conductivity are, in chemistry, chemical potentials and reaction rates.
Another physical distinction is that, contrary to the electronic case, in chemistry there is no spatial separation of components. With the assumption made in Sec. \ref{sec:CRN}, the chemical inverter is a homogeneous solution.\\
For what concerns the behavior of the two  inverters, there is a crucial  difference. In the electronic case, the steepness  of the input-output curve can be  increased  by  raising the powering voltage $\Delta V = V_+ -V_-$ (Eq. (3) in~\cite{electronic}). This is not the case for the simplified and full chemical inverters where, after a certain point, an increase in $\Delta\mu$ is not reflected by an increase in the steepness. As a matter of fact, 
the upper bounds on the nonequilibrium response found in \cite{PhysRevX.10.011066} apply to both type of chemical inverters setting a limit on their maximum steepness,
		\begin{equation}
    \begin{split}
    \left|\frac{d\mu_{O}}{d\mu_{I}}\right|_{\text{simpl}} \le \tanh\left(\frac{\Delta\mu}{4}\right) <1\\ 
       \left|\frac{d\mu_{O}}{d\mu_{I}}\right|_{\text{full}} \le 2 \tanh\left(\frac{\Delta\mu}{4}\right)<2.
\end{split}
\label{eq:steepness_inequality}
\end{equation}
The derivation of those inequalities is explained in  Appendix~\ref{app:steepness}. Their implications for the Maxwell demon will be explained in  Sec.~\ref{sec:chem_dem}.

%We emphasize that this problem would not  be circumvented with a different inverter design. In chemistry, when the mass-action law is assumed, the maximum steepness of the input-output curve is fixed by the CRN topology. In particular, the maximum value is equal to the ``number of ways" in which the input influences the output. For example, the full version of the inverter introduced in the previous Section REF cannot have a steepness greater than 2.

\section{Chemical Maxwell demon}
\label{sec:chem_dem}
In this Section we introduce and study the chemical Maxwell demon.
We start with a preliminary overview where we discuss its structure, its qualitative thermodynamics, and working principle. A detailed quantitative analysis will follow.\\

\begin{figure*}[htbp!]
\centering
\includegraphics[scale=0.3]{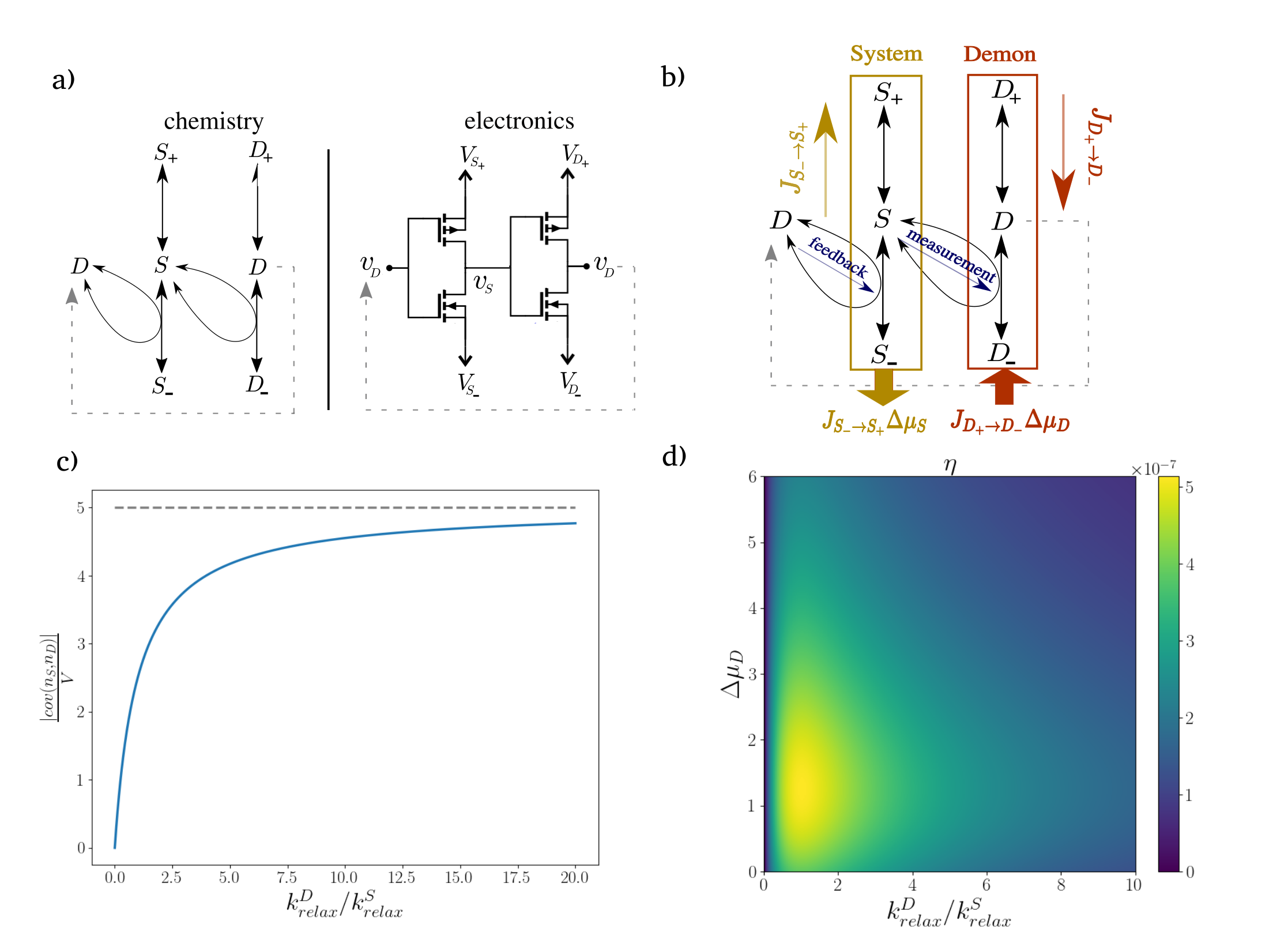}
\caption{Chemical Maxwell demon: a) Comparison of the chemical and electronic versions. (Image readapted from~\cite{electronic}). b) The CRN of the chemical Maxwell demon with its structure made explicit. c) The absolute value of the rescaled covariance as a function of the time-scale separation $k_{\text{relax}}^D/k_{\text{relax}}^S$ between the demon and the system. The dotted line represents the limiting value obtained for a fast demon $-\alpha n_S^*/V$, Eq.~\eqref{eq:correlations fast demon}. The parameters are chosen so that $\mu_{U_+} =-\mu_{U_+} =3 \,(k_BT) $ and $k^D_+ V = k_-^Dn_S^*$. d) The efficiency as a function of the time-scale separation between the demon and the system $k_{\text{relax}}^D/k_{\text{relax}}^S $ and the demon powering voltage $\Delta\mu_D$ ($\mu_{D_+}$ is kept fixed while $\mu_{D_-}$ is progressively lowered). The parameters are $n^*_S =100,\, \mu_{D_+} = 4.6 \,(k_BT),\, V k^D_+/k^D_- = V k^S_+/k^S_- = 100.$}
\label{fig:MD}
\end{figure*}

\emph{Structure.}
 As in the electronic case~\cite{electronic}, the Maxwell demon is obtained by combining two (simplified) chemical inverters.
In Fig.~\ref{fig:MD} a), the resulting MD CRN is contrasted with its electronic analogue. 
In Fig.~\ref{fig:MD} b), the CRN's structure is decomposed into its functions. One chemical inverter plays the role of the system and the other plays the role of the demon. While at the network level the demon and the system are distinct subnetworks, we notice again that  they  physically constitute the same homogeneous solution. This solution
contains six different chemical species: $S_-,S,S_+$, which belong to the system, and $D_-,D,D_+$, which pertain to the demon. The four terminal species $S_-,S_+,D_-,D_+$ are chemostatted:  their concentrations are fixed so that the upper species (Fig.~\ref{fig:MD} b)) have higher chemical potentials, $\mu_{S_+}>\mu_{S_-}$ and $\mu_{D_+}>\mu_{D_-}$. In this way, if we were to consider the system and the demon separately, we would expect the chemical reactions  to  proceed in the direction ${S_+}\to{S_-}$ and ${D_+}\to{D_-}$.\\
Since these four species are chemostatted, we are left with only two dynamical species, which are  $S$ and $D$. Therefore, the global state is specified by the numbers of molecules $\boldsymbol n =(n_S,n_D)$: $n_S$ represents the state of the system, while $n_D$  the state of the demon. The global dynamics is a  Markovian jump-process in this 2-D state space. The interaction between the system and the demon occurs through the two enzymatic reactions depicted in Fig.~\ref{fig:MD} b): $S$ is an enzyme for the demon's inverter and $D$ is an enzyme for the system's inverter. The peculiarity of enzymatic reactions is that they allow for a mutual influence between the two subnetworks \emph{without} any exchange of energy. In fact, the only thing an enzyme does is to increase the rate of the catalyzed reaction by the same amount both in the forward and backward direction. Therefore, it does not push it in either direction leaving  the reaction's equilibrium unchanged.
The first enzymatic reaction, the one catalyzed by $S$, can be interpreted as a way for the demon to collect information about the current state  of the system $n_S$ and update its state accordingly. The second enzymatic reaction, the one catalyzed by $D$, is the means by which the demon influences the system: based on its current state $n_D$, which constitutes its internal representation (knowledge) of the system, the demon outputs a feedback on it.\\

\emph{Thermodynamics.}
\label{sec:qualitative thermo}
As explained below,
through the ``measurement-feedback" mechanism just mentioned, the demon
manages under certain conditions to create a reverse current $J_{S_-\to S_+} $ in the system, Fig.~\ref{fig:MD} b). 
From  Eq.~\eqref{eq: dissipation chemostats}, this current translates into free energy being generated in the system's chemostats that can be equally seen as a local negative entropy production, 
\begin{equation}
   \dot{\mathcal{F}}_S = -T\dot \sigma_S  =J_{S_-\to S_+} \Delta \mu_S
   \label{eq:F consumption in S}
\end{equation}
with $\Delta \mu_S = \mu_{S_+} - \mu_{S_-}$. However, at the same time, the demon needs to consume free energy from its chemostats at a rate $\dot{\mathcal{F}}_D$ in order to do its job,
\begin{equation}
\dot{\mathcal{F}}_D=  -T\dot \sigma_D =  -  J_{D_+\to D_-} \Delta \mu_D 
\label{eq:F consumption in D}
\end{equation}
with $\Delta \mu_D = \mu_{D_+} - \mu_{D_-}$. 
Since, from a global perspective, the following inequality holds true~\cite{PhysRevX.4.031015},
\begin{equation}
\dot{\mathcal{F}} = \dot{\mathcal{F}}_S +\dot{\mathcal{F}}_D \le 0,
\end{equation}
the Maxwell demon as a whole is nothing but a free energy transducer that   operates in compliance with the second law and the transduction efficiency is given by the ratio 
\begin{equation}
    \eta  = \frac{\dot{\mathcal{F}}_S }{-\dot{\mathcal{F}}_D } =  \frac{J_{S_-\to S_+} \Delta \mu_S}{J_{D_+\to D_-} \Delta \mu_D}.
    \label{eq:global_efficiency}
\end{equation}
The distinctiveness of this free energy transduction is that it is solely mediated by information exchange: enzymatic reactions, as  mentioned above, do not lead to any energy exchange between the system and the demon. 
This information flow and its precise link with the thermodynamics  will be better discussed in Sec.~\ref{sec:info flow}.\\ 

\emph{Working principle.}
How does the demon manage to generate a  reverse current $J_{S_-\to S_+}$ in the system? The working principle is analogous to the electronic case: it rectifies  thermal fluctuations of chemical reactions.
To explain it, let's assume $\Delta\mu_S = 0$ for simplicity. In this condition, the system alone would reach an equilibrium with $J_{S_-\to S_+} = 0$. 
%What the demon can do, in order to generate $J_{S_-\to S_+} > 0$, , is taking advantage of fluctuations.  
The demon can achieve $J_{S_-\to S_+} > 0$ by using a particular strategy:
at steady state, any particular fluctuation $\delta n_S$ in the system's number of molecules is expected on average to decay; based on the sign of $\delta n_S$, the demon can decide how it will relax.
If $\delta n_S>0$, the demon can ``inhibit" the lower reaction by decreasing $n_D$ and making it easier for the excess of $S$ molecules to pour into  $S_+$, Fig.~\ref{fig:MD} b).  If instead $\delta n_S<0$, the demon can boost the lower reaction by increasing $n_D$ and thus promoting the replenishment of missing $S$ molecules through the conversion $S_-\to S$. In both situations, an upward current $J_{S_-\to S_+}>0$ is favored.\\
Therefore, we see that the essential ingredient  to implement this strategy is to have anticorrelations between $\delta n_S$ and $\delta n_D$.
But these anticorrelations are  guaranteed by the fact that the demon behaves as an inverter, thus,  ``inverting"  input fluctuations. 
In the above reasoning,  we are assuming that the demon is quick enough to readjust to system's fluctuations before these ones  actually decay. In  Sec. \ref{sec:computing correlations}, we will analyze the fallout from potential demon's delays.\\

In conclusion,  we see that the Maxwell demon's mechanism manifest itself in 
the correlations between the numbers of molecules. 
However, in the limit $V\to\infty$, the covariance between these numbers goes to zero in relative terms.
As a consequence, the Maxwell demon effect becomes weaker and weaker until it disappears. 
 One way to prevent this  is to increase the inverter's steepness.  A greater steepness allows for an  amplification of small fluctuations  as  will be clarified in Sec.~\ref{sec:setup}.  By rescaling it properly, in principle, one could make the output of the MD survive in the large volume limit. This is exactly what was shown for the electronic demon~\cite{electronic}. However, the same is not possible in the chemical case due to the bounds of Eq.~\eqref{eq:steepness_inequality}.\\
 To investigate how the performances of the chemical Maxwell demon scale with the system's size, we perform its analysis at steady state focussing in the  limit $V\to \infty$.
\subsection{Setup}
\label{sec:setup}
In this paragraph, we assign the rates to the Maxwell demon; we describe how the macroscopic limit is performed; we mention the parameters that will allow us to rewrite the analytical results in a compact form; and, finally, we outline the strategy adopted to analytically solve the Maxwell demon.\\
The rate assignment is analogous to expression~\eqref{eq:rate choice simplified inverter} both for the system and demon inverters:
\begin{equation}
    \ce{ $S_-$ <=>[$ k_-^S \boldsymbol{{\color{BrickRed}n_D} }\exp(\mu_{S_-})$][$k_-^S n_S \boldsymbol{{\color{BrickRed}n_D}}/V$] $S$ <=>[$k_+^S n_S$][$ k_+^S V\exp(\mu_{S_+})$] $S_+ $  }
\label{eq:rate choice S}\end{equation}
\begin{equation}
    \ce{ $D_-$ <=>[$ k_-^D\boldsymbol{{\color{olive} n_S}} \exp(\mu_{D_-})$][$k_-^D \boldsymbol{{\color{olive} n_S}} n_D/V$] $D$ <=>[$k_+^D n_D$][$ k_+^D V\exp(\mu_{D_+})$] $D_+ $  }.
\label{eq:rate choice D}\end{equation}
In red and yellow, the influences of the enzymes $S$ and $D$ on the rates are highlighted. 
 As in Eq.~\eqref{eq: chem pot without standard}, we  measure$\mu_{S_+},\mu_{S},\mu_{S_-}$ with respect to $\mu_S^0$ and $\mu_{D_+},\mu_{D},\mu_{D_-}$  with respect to $\mu_D^0$  to filter out the standard chemical potentials of the $S$ and $D$ species. This rate assignment is in compliance with the local detailed balance condition \eqref{eq:ldb}, thus ensuring the thermodynamic consistency of the Maxwell demon's model.
\\
We denote with $x$ and $y$ respectively the system and demon's concentrations, 
\begin{equation}
    x= \frac{n_S}{V}\quad y =  \frac{n_D}{V}.
\end{equation}
The macroscopic limit is taken in the following manner: 
\begin{equation}
    \begin{split}
        \mu_{D_-},\mu_{D_+}, \mu_{S_-} = \text{fixed}\\
        \mu_{S_+}\to  \mu_{S_-}  \quad\text{as} \quad V\to\infty
    \end{split}
\end{equation}
The precise scaling of  
$\Delta \mu_S$ will be later derived, Eq.~\eqref{eq:affinity correlations}.
 The reason why $\Delta \mu_S\to 0$ is that the Maxwell demon effect becomes weaker and weaker, thus allowing to overcome decreasing chemical gradients $\Delta \mu_S$. In the limit $V\to \infty$, the deterministic concentrations of the $S$ and $D$ species are, from Eq.~\eqref{eq:i-o relation}:
     \begin{equation}
      (x^*,y^*)  =\left(e^{\mu_{S_-}},\frac{k_{+}^D \: e^{\mu_{D_+}}+k_{-}^D \: e^{\mu_{S_-}+\mu_{D_-}} } {k_{+}^D+k_{-}^D \: e^{\mu_{S_-}}}\right).
        \label{eq:working point}
 \end{equation}
 We denote $(x^*,y^*)$ as  the working point.
\\
In the following, three quantities will turn out to be particularly useful for interpreting the analytical results: the two inverter relaxation rates $k_{\text{relax}}^S ,k_{\text{relax}}^D $ and the fluctuation-amplifying factor $\alpha$. 
The inverter relaxation rates are approximately given by Eq.~\eqref{eq:relax_rate} evaluated in the macroscopic limit 
\begin{equation}
    \begin{split}
        k_{\text{relax}}^S = k_{+}^S+k_{-}^Sy^*\\
        k_{\text{relax}}^D = k_{+}^D+k_{-}^Dx^*.
    \end{split}
\end{equation}
The fluctuation-amplifying factor $\alpha$ can be  introduced through the function $n_D(n_S)$ that corresponds to the steady state input-output relation of the demon inverter rewritten in terms of number of molecules. In other words, $n_D(n_S)$ is the average number  of $D$ molecules assuming that the number of $S$ molecules is kept fixed at the value $n_S$. From Eq.~\eqref{eq:i-o relation}, it is
\begin{equation}
    \frac{ n_D(n_S)}{V}= y =\left[\frac{k_{+}^D \: e^{\mu_{D+}}+k_{-}^D \: e^{\mu_{D-}} \: x} {k_{+}^D+k_{-}^D \:x}\right].
    \label{eq:demon_input_output}
\end{equation}
The fluctuation-amplifying factor is then defined as
\begin{equation}
    \alpha = -  \frac{d n_D }{dn_S} (n_S^*) = \frac{k^D_+k^D_-\left(e^{\mu_{D_+}}-e^{\mu_{D_-}}\right)}{(k^D_{\text{relax}})^2}.
    \label{eq:alpha def}
\end{equation}
Put into words,
$\alpha>0$ is the factor by which small fluctuations $\delta x = \delta n_S/V$ around the working point get amplified by the demon's inverter assuming it reacted instantaneously. In principle, we would like to increase $\alpha$ as much as possible: the bigger it is, the stronger the Maxwell demon effect. However, 
the  inequality for the steepness of the input-output relation~\eqref{eq:steepness_inequality}  sets an upper bound to its  value:
\begin{equation}
    \alpha  \le \frac{n_D^*}{n_S^*} \tanh\left(\frac{\Delta\mu_D}{4}\right) <\frac{n_D^*}{n_S^*}
    \label{eq:alpha inequality}.
\end{equation} 
We now present the strategy adopted to solve the Maxwell demon. At steady state, one must have for each inverter equal upper and lower average currents, $J_{S_+\to S} =  J_{S\to S_-}$ and $J_{D_+\to D} =  J_{D\to D_-}$ (see Fig.\ref{fig:MD} b)), and therefore:
\begin{equation}
     k_+^S(V e^{\mu_{S_+}} -\langle n_S\rangle)
     = k_-^S\left(\frac{\langle n_S n_D\rangle}{V} -\langle n_D\rangle \: e^{\mu_{S_-}} \right), 
\label{eq:steady state condition 1}
\end{equation}
and
\begin{equation}
   k_+^D(V e^{\mu_{D_+}} -\langle n_D\rangle)  = k_-^D\left(\frac{\langle n_S n_D\rangle}{V} -\langle n_S\rangle \: e^{\mu_{D_-}} \right). 
\label{eq:steady state condition 2}
\end{equation}
From the left-hand side of the above expressions, we see that $J_{S_-\to S_+}$, $J_{D_+\to D_-}$ and thus the efficiency $\eta$, Eq.~\eqref{eq:global_efficiency}, can be obtained once we know $\langle n_S\rangle$ and $\langle n_D\rangle$. In particular, 
since $J_{D_+\to D_-}$ scales $\propto V$, it can be calculated approximating $\langle n_D\rangle \simeq n_D^*$. Using Eq.\eqref{eq:working point} for $n_D^*$ and substituting inside Eq.\eqref{eq:steady state condition 2} leads to
\begin{equation}
    J_{D_+\to D_-} \simeq \frac{k_{+}^D k_{-}^D n_S^*\left(e^{\mu_{D_+}}-e^{\mu_{D_-}}\right)} {k_{\text{relax}}^D} = V\alpha \,k^D_{\text{relax}}x^*.
    \label{eq:current demon}
\end{equation}
Notice that $J_{D_+\to D_-}$
and thus the free energy consumed by the demon  is  also $\propto\alpha$, the fluctuation-amplifying factor.
To calculate $J_{S_-\to S_+}$, we need to be more careful as it has a much smaller value: differently from $J_{D_-\to D_+}$, it turns out to be intensive in the volume.
\begin{equation}
    J_{S_-\to S_+}(V) = \cancel{J_{S_-\to S_+}^{(0)}} V + J_{S_-\to S_+}^{(1)} + \frac{J_{S_-\to S_+}^{(2)}}{V} + \dots
\end{equation}
Therefore, the same approximation $\langle n_S\rangle \simeq n_S^*$ that would be right to
obtain the zeroth order of the current fails to give a correct result for $J_{S_-\to S_+}^{(1)}$.
To obtain this value, we would need to calculate $\langle n_S\rangle$ with higher accuracy, up to order $O(1)$.
The procedure we will follow involves two steps: first, we show that we can link $J_{S_-\to S_+}$ and thus all the quantities characterizing the Maxwell demon performances to the  covariance $\text{cov}(n_S,n_D) \equiv \langle n_S n_D \rangle - \langle n_S \rangle\langle n_D \rangle$; secondly, we calculate the latter through a Gaussian approximation of the rate function. \\
 Note that a more straightforward approach in which one uses the Gaussian approximation to directly calculate $\langle n_S\rangle$ is wrong. In general, the Gaussian approximation is not accurate
enough for first order moments, while it is for second order moments, like $\text{cov}(n_S,n_D)$. See Appendix~\ref{app: accuracy rate function} for a discussion. \\

\subsection{ Maxwell demon's quantities  as a function of $\text{cov}(n_S,n_D)$}
\label{sec:linking correlations}
The covariance $\text{cov}(n_S,n_D)$ emerges as a pivotal variable. In this paragraph, we link  $J_{S_-\to S_+}$ and  all the other quantities describing the Maxwell demon  performances to it. %that is, the correlations generated by its operation. 
This allows us to address the scaling behavior of those quantities from the expected scaling 
\begin{equation}
    \text{cov}(n_S,n_D)\propto V.
    \label{eq: corr scaling}
\end{equation}
as the covariance is additive for independent systems. This scaling will be later confirmed in Eq. \eqref{eq: correlations}.
% In the next paragraph, we obtain a formula  for the correlations.\\ 
 Firstly, we establish the relationship between $J_{S_-\to S_+}$  and $\text{cov}(n_S,n_D)$. We do it through Eq.~\eqref{eq:steady state condition 1} working up to first order in $\langle x\rangle - x^*$ and $\langle y\rangle - y^*$. Calculations are reported in  Appendix~\ref{app: connection with corr}; the result is
     \begin{equation}
   J_{S_-\to S_+}= k^S_{\text{relax}}n_S^* \left( -\frac{\text{cov}(n_S,n_D)}{n_S^*n_D^*}  -\Delta\mu_S\right) \!\left[\frac{k_+^S k_-^S y^*}{(k^S_{\text{relax}})^2}\right]\!. 
   \label{eq:current system correlation}
\end{equation}
The last factor on the right-hand side is adimensional and only depends on the system's rates. Interestingly, the reverse current generated by the Maxwell demon is proportional to $\propto -\text{cov}(n_S,n_D)>0$, which is partly due to the fact that the interaction between $S$ and $D$ occurs through \emph{two-body} reactions (i.e. the enzymatic reactions). Because of Eq.~\eqref{eq: corr scaling}, we see that  $J_{S_-\to S_+}$ is intensive, it does not vary with the system's size. From expression~\eqref{eq:current system correlation}, we also see that the effective reverse affinity created in the system is equal to 
\begin{equation}\mathcal{A}_{\text{eff}} = -\frac{\text{cov}(n_S,n_D)}{n_S^*n_D^*} \propto \frac{1}{V}.
    \label{eq:affinity correlations}
\end{equation}
$0<\Delta\mu_S<\mathcal{A}_{\text{eff}}$ represents the range of vincible chemical gradients. It goes to zero in the deterministic limit.  \\
Having connected the current $J_{S_-\to S_+}$  to the covariance $\text{cov}(n_S,n_D)$, we can do the same for the output power $\dot{\mathcal{F}}_S$, Eq.~\eqref{eq:F consumption in S}, and the efficiency $\eta$, Eq.~\eqref{eq:global_efficiency}. One obtains
    \begin{equation}
\dot{\mathcal{F}}_S =  k_{\text{relax}}^S n_S^*\Delta\mu_S  
    \left( -\frac{\text{cov}(n_S,n_D)}{n_S^*n_D^*}  -\Delta\mu_S\right) \left[\frac{k_+^S k_-^S y^*}{(k^S_{\text{relax}})^2}\right],
\end{equation}
and
\begin{equation}
    \eta =
    \frac{\left( -\frac{\text{cov}(n_S,n_D)}{n_S^*n_D^*}  -\Delta\mu_S\right) \Delta\mu_S}{\alpha\Delta\mu_D} \left( \frac{k_{\text{relax}}^S}{k_{\text{relax}}^D}\right) 
    \left[\frac{k_+^S k_-^S  y^*}{(k^S_{\text{relax}})^2}\right]. 
\end{equation}
The efficiency and the output power are quadratic functions of $\Delta\mu_S$ and they are both maximized by 
\begin{equation}
    \Delta\mu_S^{\text{max}} =- \frac{1}{2}\frac{\text{cov}(n_S,n_D)}{n_S^*n_D^*}.  
    \label{eq:max delta mu S}
\end{equation}
This optimal value leads to
    \begin{equation}
\max_{\Delta\mu_S} \dot{\mathcal{F}}_S =\frac{k_{\text{relax}}^S n_S^*}{4}
    \left(\frac{\text{cov}(n_S,n_D)}{n_S^*n_D^*}\right)^2 
 \left[\frac{k_+^S k_-^S  y^*}{(k^S_{\text{relax}})^2}\right], 
\end{equation}
and
\begin{equation}
    \max_{\Delta\mu_S} \eta = \frac{1}{4\alpha\Delta\mu_D}
    \left(\frac{\text{cov}(n_S,n_D)}{n_S^*n_D^*}\right)^2 
   \! \left( \frac{k_{\text{relax}}^S}{k_{\text{relax}}^D}\right) \!\left[\frac{k_+^S k_-^S  y^*}{(k^S_{\text{relax}})^2}\right]. 
    \label{eq: max efficiency correlations}
\end{equation}
The optimal output power and efficiency turn out to be proportional to the square of the covariance and their scaling, from Eq.~\eqref{eq: corr scaling}, is respectively
$\dot{\mathcal{F}}_S \propto 1/V$ and $\eta \propto 1/V^2$. This is due to the fact that the opposite chemical gradient 
$\Delta\mu_S^{\text{max}}$ is $\propto V^{-1}$ plus, for the efficiency, the fact that the current in the demon's inverter  $J_{D_+\to D_-}$ is $ \propto V$.
In Table \ref{tab:scaling}, we summarize these scalings characterizing the Maxwell demon; they are analogous to the ones found for the electronic version~\cite{electronic,electronic_info_flow}.
\renewcommand{\arraystretch}{1.5} %
\begin{table}[!h]
    \centering
\begin{tabular}{ |c|c| } 
\hline
Quantity & Scaling \\
\hline 
$J_{D_+\to D_-}$, cov($n_S$,$n_D$),  $\dot{\mathcal{F}}_D$   & $\propto V$ \\ 
$J_{S_-\to S_+}$, $\dot{\mathcal{I}}$ & intensive \\ 
$\Delta\mu_S^{\text{max}}$, $\dot{\mathcal{F}}_S $, $\eta_S$, $\eta_D$ & $\propto \frac 1V$\\
$\eta $& $\propto \frac{1}{V^2}$\\
\hline
\end{tabular}
\caption{Summary of the MD's behavior in the limit $V\to\infty$. We report the
scaling of: the system and demon's currents $J_{S_-\to S_+}$, $J_{D_+\to D_-}$; the covariance cov($n_S$,$n_D$); the input and output power $\dot{\mathcal{F}}_D$, $\dot{\mathcal{F}}_S$; the optimal adverse chemical gradient $\Delta\mu_S^{\text{max}}$;  the global and partial efficiencies $\eta $, $\eta_S$, $\eta_D$; and the information flow $\dot{\mathcal{I}}$.}
    \label{tab:scaling}
\end{table}
 \subsection{ $\text{cov}(n_S,n_D)$}
 \label{sec:computing correlations}
As explained in  Appendix~\ref{app: accuracy rate function}, $\text{cov}(n_S,n_D)$  can be determined with a sufficient accuracy from the 
Gaussian approximation of the rate function. This approximation can be derived, as was done in~\cite{electronic}, from the master equation of the 2D jump-process. The schematic procedure consists in, firstly, substituting 
\begin{equation}
    P_{\rm ss}(\boldsymbol n)\asymp e^{-\frac{1}{2 V} (\boldsymbol{n} -\boldsymbol{n}^*) C^{-1} (\boldsymbol{n} -\boldsymbol{n}^*) }
    \end{equation}
in the master equation and, secondly, solving the resulting equation for $C$, the covariance matrix. The details are reported in Appendix~\ref{app: evaluation of correlations}. The final result is
\begin{equation}
    \text{cov}(n_S,n_D) = -\alpha n_S^*\left(\frac{k_{\text{relax}}^D/k_{\text{relax}}^S}{k_{\text{relax}}^D/k_{\text{relax}}^S+1 }\right).
    \label{eq: correlations}
\end{equation}
 In Fig.~\ref{fig:MD} c), we plot it as a function of the time-scale separation $k_{\text{relax}}^D/k_{\text{relax}}^S $ between the demon and the system.\\
 When $k_{\text{relax}}^D/k_{\text{relax}}^S\to \infty$, we are in the fast demon limit: the demon inverter responds instantaneously to changes in $n_S$ according to Eq.~\eqref{eq:demon_input_output}. Since fluctuations $ \delta x = \delta n_S/V$ around the working point are small when $V\to\infty$, the demon's response can be actually approximated with the linear part of Eq.~\eqref{eq:demon_input_output}:
 \begin{equation}
     n_D(n_S) - n_D^* \simeq -\alpha (n_S - n_S^*).
     \label{eq:linear relation}
 \end{equation}
and the covariance between $n_S$ and $n_D$ arising from such a response is \footnote{We use the fact that, in the absence of the small perturbation introduced by the demon feedback, $P_{\rm ss}(n_S)$ would be a Poissonian with average $\simeq n_S^*$.}
 \begin{equation}
     \text{cov}(n_S,n_D) = -\alpha \langle(n_S-n_S^*)^2\rangle \simeq -\alpha \langle \Delta n_S^2\rangle \simeq -\alpha n^*_S,
\label{eq:correlations fast demon}
 \end{equation}
which is in agreement with the value given by Eq.~\eqref{eq: correlations}.\\
On the other hand, when $k_{\text{relax}}^D/k_{\text{relax}}^S\to 0$, the demon becomes very slow and its delay results in weaker correlations:
$D$ tries to update its state based on that of $S$, but since it is slow, $S$ may change in the meantime. This is confirmed from Eq.~\eqref{eq: correlations}
\begin{equation}
    \text{cov}(n_S,n_D) \propto k_{\text{relax}}^D/k_{\text{relax}}^S\to 0.
\end{equation}
From Eq.~\eqref{eq:current system correlation} 
 and \eqref{eq:affinity correlations}, the weaker the correlations, the lower the reverse current $J_{S_-\to S_+}$ and the affinity $\mathcal{A}_{\text{eff}}$  produced in the system. In other words, 
this delay makes the demon feedback on the system anachronistic and thus less effective.

\subsection{Efficiency $\eta$}
\label{sec:efficiency}
With the covariance calculated in the previous paragraph, we can obtain an explicit expression for the efficiency of Eq.~\eqref{eq: max efficiency correlations}. The result is
\begin{equation}
    \max_{\Delta\mu_S} \eta = \frac{\alpha}{4V^2 (y^*)^2\Delta\mu_D }  \frac{k_{\text{relax}}^D/k_{\text{relax}}^S}{(k_{\text{relax}}^D/k_{\text{relax}}^S+1)^2 }
  \left[\frac{k_+^S k_-^S  y^*}{(k^S_{\text{relax}})^2}\right].
    \label{eq: max efficiency delta mu S}
\end{equation}
A plot of $\eta$ as a function of $\Delta\mu_D$ and the time-scale separation $k_{\text{relax}}^D/k_{\text{relax}}^S $ between the demon and the system is shown in Fig.~\ref{fig:MD} d).
When $k_{\text{relax}}^D/k_{\text{relax}}^S \to 0$, $\eta\to 0$
because of the negative impact of the demon's delay. Also, when $k_{\text{relax}}^D/k_{\text{relax}}^S \to \infty$, $\eta\to 0$: the demon answers promptly to any system's change but, in order to do so, it  consumes an increasing amount of free energy $\propto  J_{D_+\to D_-} \propto k_{\text{relax}}^D$. 
From Eq.~\eqref{eq: max efficiency delta mu S}, the efficiency is maximized when the time-scales of the two inverters are equal: $k_{\text{relax}}^D=k_{\text{relax}}^S$.  When $\Delta\mu_D\to \infty$, $\eta \to 0$:  after a certain point, an increase in the demon powering voltage does not translate into an increase of
$\frac{\alpha}{n_D^*}$ (Eq.~\eqref{eq:alpha inequality}) and leads only to more free energy being dissipated. From the plot, we see that the efficiency is maximized for a nonzero $\Delta\mu_D$. Its maximum value can be upper-bounded exploiting the inequality~\eqref{eq:alpha inequality} and the fact that the right-most adimensional parentheses of Eq.~\eqref{eq: max efficiency delta mu S} is $\le 1/4$.
\begin{equation}
    \eta \le \frac{1}{256 n_D^*n_S^* }.
   \label{eq: max efficiency }
\end{equation}
%In adiabatic regime, the efficiency becomes
%\begin{footnotesize}
 %   \begin{equation}
  %%  \max \eta = \frac{1}{16 n_D^*n_S^* }  %\left[\frac{k_+^D k_-^Dn_S^*/V}%{(k^D_{\text{relax}})^2}\right] 
  %\left[\frac{k_+^S k_-^S n_D^*/V}{(k^S_{\text{relax}})^2}\right] \le \frac{1}{256 n_D^*n_S^* }
  %  \label{eq: max efficiency }
%\end{equation}
%\end{footnotesize}
The maximum value is very low as it is inversely proportional to the product of the  numbers of $S$ and $D$
molecules present in the solution. For example,   in biological conditions, imagining $S$ and $D$ to be proteins inside a cell with $n_S^*,n_D^*\simeq 10^2-10^3$, we would have $\eta< 10^{-6}-10^{-8}$.

\subsection{Maxwell demon thermodynamics with information flow}
\label{sec:info flow}
In Section~\ref{sec:qualitative thermo}, we presented the basic Maxwell demon
thermodynamics and we highlighted the absence of any energy exchange between the two inverters. 
In this Section, we dive deeper analyzing the Maxwell demon with the formalism of bipartite systems~\cite{PhysRevX.4.031015,article}.
This framework allows one to further resolve
its thermodynamics. In particular,  two \emph{separate} second laws can be obtained for the two inverters. They include a new term, the information flow, that quantifies the rate at which information is exchanged between the system and the demon. Thanks to these refined second laws, it is possible to  decompose the global efficiency, Eq.~\eqref{eq:global_efficiency}, as the
product of two efficiencies related respectively to the demon and the system, or equivalently, to the measurement and feedback processes. In this section, we first explain how the bipartite formalism applies to our chemical MD and then evaluate the information flow, which allows us to discuss the behavior of the two efficiencies above mentioned.\\
We begin by stating in which sense the Maxwell demon is  bipartite: it is composed of two degrees of freedom, which are the state of the system $n_S$ and the state of the demon $n_D$; in addition,  the possible chemical reactions change either $n_S$ or $n_D$ but not both at the same time.  This last property  enables a decomposition of the time derivative of the mutual information 
\begin{equation}
    \mathcal{I} =\sum_{n_S,n_D} P(n_S,n_D) \log\frac{ P(n_S,n_D)}{ P(n_S)P(n_D)}
\end{equation}
into two pieces: one due to reactions changing the state of the system and the other due to reactions changing the state of demon:
\begin{equation}
    d_t \mathcal{I} =  \dot{\mathcal{I}}_S + \dot{\mathcal{I}}_D,
\end{equation}
with 
\begin{equation}
\begin{split}
    \dot{\mathcal{I}}_S  =\sum_{n_D} \sum_{n_S'>n_S} j(n_S'\leftarrow n_S|n_D)\log \frac{P(n_D|n_S')}{P(n_D|n_S)}\\
    \dot{\mathcal{I}}_D  = \sum_{n_S} \sum_{n_D'>n_D} j(n_D'\leftarrow n_D|n_S)\log \frac{P(n_S|n_D')}{P(n_S|n_D)}.
    \end{split}
    \label{eq:info flow expressions}
\end{equation}
In these expressions, $j(n_S'\leftarrow n_S|n_D)$ represents the net average current along  the transition $(n_S,n_D)\to (n_S',n_D)$ and $j(n_D'\leftarrow n_D|n_S)$ is defined analogously.
In stationary conditions, one must have 
\begin{equation}
    d_t \mathcal{I} = 0\implies  \dot{\mathcal{I}}_D = - \dot{\mathcal{I}}_S =  \dot{\mathcal{I}}.
\end{equation}
We call $\dot{\mathcal{I}}$ the information flow from the demon to the system. 
As anticipated, for bipartite systems, it is possible to derive two refined second laws valid for each subpart that take into account this quantity~\cite{PhysRevX.4.031015}. They read
\begin{equation}
    \begin{split}
        \dot\sigma_S + k_B \dot{\mathcal{I}}\ge 0\\
              \dot\sigma_D - k_B \dot{\mathcal{I}}\ge 0,
    \end{split}
    \label{eq:refined_2law}
\end{equation}
where $\dot\sigma_{S/D}$ is the steady state local dissipation rate of the system/demon inverter defined in Eq.~\eqref{eq:F consumption in S}/\eqref{eq:F consumption in D}. These two inequalities embody 
the thermodynamic understanding of a Maxwell demon: the demon produces mutual information (correlations), $\dot{\mathcal{I}}>0$, by adjusting its state according to that of the system. However, to do this, it needs to dissipate a certain amount of free energy $\dot\sigma_D\ge k_b\dot{\mathcal{I}} $. This generated mutual information is subsequently burned by the system allowing for a negative local dissipation $-k_B\dot{\mathcal{I}}\le\dot\sigma_{S}<0$, which corresponds to free energy being extracted from the system. The sum of both inequalities of Eq.~\eqref{eq:refined_2law}, 
\begin{equation}
    \dot\sigma_{S} + \dot\sigma_{D}\ge 0,
\end{equation}
ensures that the Maxwell demon as a whole operates in compliance with the second law. \\
The refined second laws, Eq.~\eqref{eq:refined_2law}, also allow us to decompose the overall transduction efficiency, Eq.~\eqref{eq:global_efficiency}, into two partial efficiencies:
\begin{equation}
    \eta  = \eta_S\, \eta_D = \left(\frac{J_{S_-\to S_+} \Delta \mu_S}{ \dot{\mathcal{I}}}\right)\left(\frac{ \dot{\mathcal{I}}}{J_{D_+\to D_-} \Delta \mu_D}\right).
    \label{eq:global_efficiency_splitting}
\end{equation}
The partial efficiency $\eta_D$ measures how efficiently the demon converts the consumed free energy into new mutual information, while $\eta_S$ is the efficiency with which the system burns mutual information to release free energy in its chemostats. $\eta_D$ and $\eta_S$ can be, respectively, thought of as the efficiencies of the measurement and feedback step, and are less or equal to one, by virtue of the inequalities in Eq. \eqref{eq:refined_2law}.
\\
 \begin{figure*}[htbp!]
\centering
\includegraphics[scale=0.31]{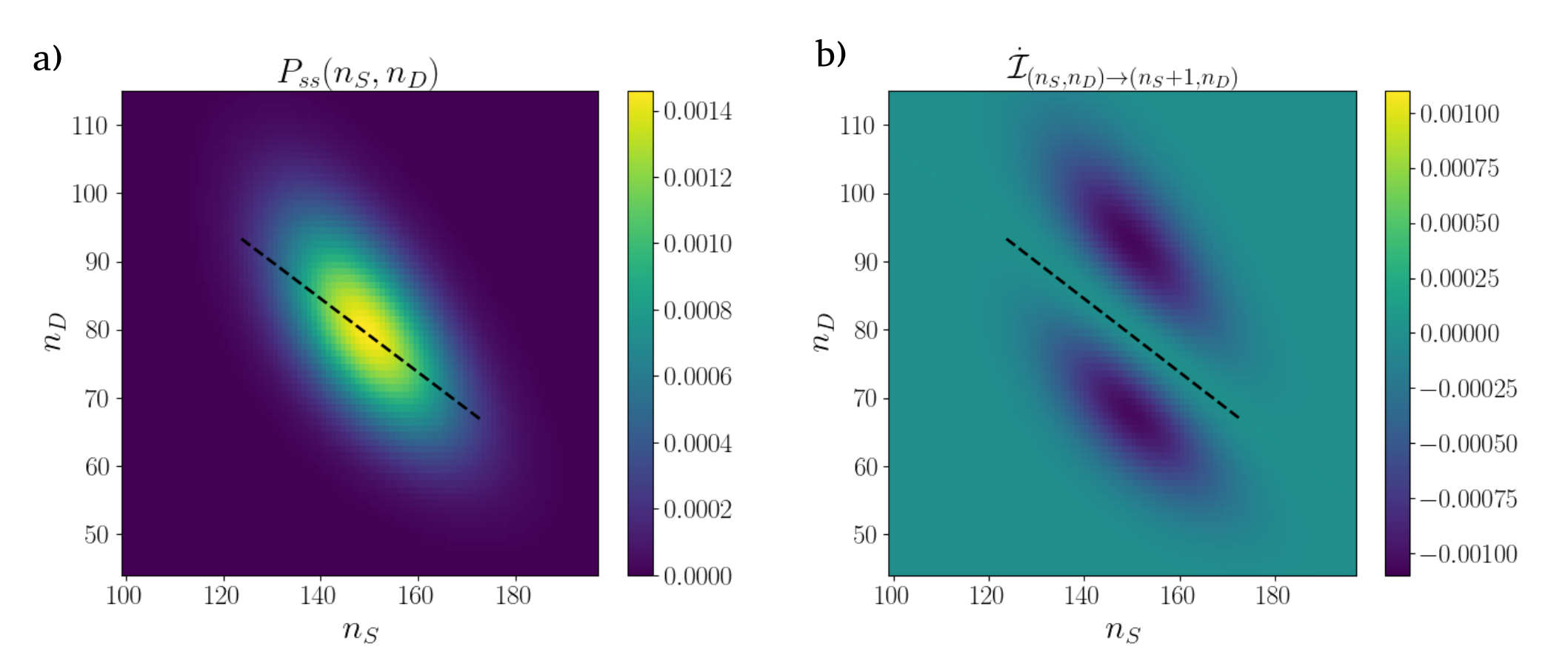}
\caption{Information flow's pattern in the state space (fast demon limit): a) stationary probability distribution; b) in point $(n_S,n_D)$ it is shown the individual contribution to the information flow coming from the horizontal transition ${(n_S,n_D)\to(n_S+1,n_D)}$. In both pictures, the dotted black line represents the approximated  linear demon's response $n_D(n_S)$, Eq.~\eqref{eq:linear relation}. The parameters are $n_S^* = 148,\, n_D^* = 80,\, \alpha  = n_D^*/n_S^*,\, V k_+^S/k_-^S = 100$.}
\label{fig:info_flow}
\end{figure*}
\paragraph{$\dot{\mathcal{I}}$ in the chemical Maxwell demon}
 We derive $\dot{\mathcal{I}}$ analytically for a fast demon and we find its qualitative behavior in the case of a slow demon. Finally, we display the information flow pattern in the state space. \\
In the limit $k_{\text{relax}}^D\gg k_{\text{relax}}^S$, it is simpler to derive  $\dot{\mathcal{I}}$.  For the sake of its derivation, we can also assume $\Delta\mu_S=0$. As a matter of fact, $\dot{\mathcal{I}}$ can be considered with good approximation  constant over the small range of vincible chemical potential gradients $0\le \Delta\mu_S\le \mathcal{A}_{\text{eff}}\propto 1/V$, Eq.~\eqref{eq:affinity correlations}.
We compute $\dot{\mathcal{I}}$ from $\dot{\mathcal{I}}_S$, Eq.~\eqref{eq:info flow expressions}.  For this,  the knowledge of the steady state conditional probability distribution $P_{\rm ss}(n_D|n_S)$ is required.
In the  fast demon limit, this distribution can be approximated with a Poissonian (see the end of Sec. \ref{subsubsec:simplified_chem_inv}) with average $n_D(n_S)$ given by Eq.~\eqref{eq:demon_input_output}.   The detailed calculations are reported in Appendix~\ref{app:evaluation of I}. The result is
\begin{equation}
    \dot{\mathcal{I}} = \alpha^2 \frac{n_S^*}{n_D^*} \,k^S_{\text{relax}}.
    \label{eq:info flow fast demon}
\end{equation}
We notice that $\dot{\mathcal{I}}$ is an intensive quantity: 
it does not scale with the volume $V$.
We  can combine its expressions with that of
$J_{S_-\to S_+}$, Eq.~\eqref{eq:current system correlation},   
to obtain the partial efficiency $\eta_S$ in Eq.~\eqref{eq:global_efficiency_splitting}. This  partial efficiency is maximum for the same $\Delta\mu_S^{\text{max}}$ that maximized $\eta$, Eq.~\eqref{eq:max delta mu S}.
\begin{equation}
    \max_{\Delta\mu_S} \eta_S = \frac{1}{4n_D^*}\left[\frac{k_+^S k_-^Sy^*}{\left(k^S_{\text{relax}}\right)^2}\right]\le \frac{1}{16n_D^*}.
\end{equation}
From the above expression, we see that the partial efficiency with which the system  converts mutual information into output free energy scales $\propto 1/V$. Knowing from   Eq.~\eqref{eq: max efficiency correlations} that  $\eta\propto 1/V^2$, we conclude that $\eta_D\propto 1/V$ (see Table \ref{tab:scaling} for a summary). \\
To make the discussion more exhaustive, we can qualitatively analyze  the opposite  case of a very slow demon
 $k_{\text{relax}}^D\ll k_{\text{relax}}^S$. In this limit, we know through Eq.~\eqref{eq: max efficiency delta mu S} that $\eta \propto k_{\text{relax}}^D/ k_{\text{relax}}^S \to 0$ and
 we can ask ourselves whether this inefficiency is due to $\eta_S$ or $\eta_D$. To answer this question, it is necessary to know how the information flow scales. In  Appendix~\ref{app: scaling I}, we show that its scaling is
\begin{equation}
    \dot{\mathcal{I}} \propto  k_{\text{relax}}^D\left(\frac{k_{\text{relax}}^D}{k_{\text{relax}}^S}\right) \quad \text{for }\quad \frac{k_{\text{relax}}^D}{ k_{\text{relax}}^S}\to 0.
\end{equation}
Therefore, in the limit of a slow demon, keeping in mind that $J_{S_-\to S_+}\propto k_{\text{relax}}^S$ and $J_{D_+\to D_-}\propto k_{\text{relax}}^D$,  the partial efficiencies scale as 
\begin{equation}
    \eta_S = \text{const.}\quad  \eta_D \propto \frac{k_{\text{relax}}^D}{k_{\text{relax}}^S}\to 0.
\end{equation}
This means that the loss in efficiency is due to the measuring step: as the demon becomes slower, due to delays, it converts much less effectively  consumed free energy  $ J_{D_+\to D_-}\Delta\mu_D$ into mutual information. \\
We conclude the section by discussing the information flow pattern in the state space. As a matter of fact, from Eq.~\eqref{eq:info flow expressions}, we see that we can break down $\dot{\mathcal{I}}$ into a sum of contributions coming from each individual transition
\begin{equation}
    \dot{\mathcal{I}} =   - \dot{\mathcal{I}}_S  = -\sum_{n_S,n_D} \dot{\mathcal{I}}_{(n_S,n_D)\to(n_S+1,n_D)}.
\end{equation}
We can then visualize these contributions by plotting them in the state space: in point $(n_S,n_D)$ we show the value  $\dot{\mathcal{I}}_{(n_S,n_D)\to(n_S+1,n_D)}$. The  plot, together with that of $P_{\rm ss}(n_S,n_D)$, is shown in Fig.~\ref{fig:info_flow}. The probability distribution is a rotated ellipsoid, denoting correlations between $n_S$ and $n_D$. $\dot{\mathcal{I}}_{(n_S,n_D)\to(n_S+1,n_D)}$ is everywhere negative, meaning that the system is consuming mutual information.
The main contributions to $\dot{\mathcal{I}}_S$ arise from the two regions close to the center that are adjacent to the line of Eq.~\eqref{eq:linear relation}. We can elucidate the pattern in the following way. 
States starting from the line, where $P_{\rm ss}(n_S,n_D)$ is concentrated (Fig.~\ref{fig:info_flow} a)), diffuse horizontally to the left/right of the dotted line due to random transitions in the system's inverter. This horizontal dilution has the effect of destroying correlations between $S$ and $D$
 and, consequently, of reducing mutual information. 
 However, the demon counteracts this horizontal spreading by 
vertically absorbing back into the line those escaping states. In this way, the demon is generating in those two regions above/below the line new mutual information that compensates the one burned by the system in the same two regions.\\

\section{Conclusions}

In this article, we studied the chemical analog of an electronic Maxwell demon. We examined in detail its structure and working principle (Sec.~\ref{sec:chem_dem}). 
We highlighted the correlations between the numbers of molecules as the central quantity that captures the essence of the MD (Sect.~\ref{sec:linking correlations}). 
 We conducted a thorough characterization of the system's scaling in the macroscopic limit, summarized in Table~\ref{tab:scaling}, and examined the Maxwell demon's efficiency in relation to the time-scale separation of the two inverters, the demon powering chemical potential, and the opposing chemical gradient (Sec.\ref{sec:efficiency}).
Finally, the bipartite formalism allowed us to enlighten  the details of its thermodynamics through the information flow (Sec.~\ref{sec:info flow}).\\ 
The system presents some inconveniences:
its efficiency is found to be extremely low (Eq.~\ref{eq: max efficiency }) and it is not possible  to make the MD's output survive in the macroscopic limit even by increasing the input power.
Nevertheless, this system enriches 
our understanding of information-mediated transduction in CRNs and represents a first step towards unveiling deeper connections between electronics and chemistry. Regarding the latter point, we believe it worthwhile to push further the two approaches followed in this paper: the electronics-inspired design of CRNs and their analysis from a modular perspective \cite{avanzini} analogous to the one developed for electronic circuits.

\section{Acknowledgements}
We thank Francesco Avanzini and Emanuele Penocchio for useful discussions.
This is research was supported by AFR PhD grant 15749869 and by project ChemComplex
(C21/MS/16356329) funded by Luxembourg National
Research Fund (FNR) (Luxembourg), and by project
no. INTER/FNRS/20/15074473 funded by FRS-FNRS
(Belgium) and FNR (Luxembourg).
For open access, the author has applied a Creative Commons Attribution 4.0 International (CC BY 4.0) license to any Author Accepted Manuscript version arising from this submission.\\

\appendix
\section{Bounds on the chemical inverter steepness}
\label{app:steepness}
The first bound in~\eqref{eq:steepness_inequality} can be shown by  computing $\left|\frac{d\mu_{O}}{d\mu_{I}}\right|$  from Eq.~\eqref{eq:i-o relation}
\begin{equation}
\begin{split}
    \left|\frac{d\mu_{O}}{d\mu_{I}}\right|_{\text{simpl}} &= \frac{k_-k_+e^{\mu_I}\left(e^{\Delta\mu}-1\right)}{\left(k_+ + k_-e^{\mu_I}\right)\left(k_+e^{\Delta\mu} + k_-e^{\mu_I}\right)}\\ & \le \tanh\left(\frac{\Delta\mu}{4}\right) .
    \label{eq:first ineq}
\end{split}
\end{equation}
The last inequality can be verified directly through calculations. 
Alternatively, one can recognize that the inverter falls in the range of application of~\cite{PhysRevX.10.011066} (Sec. VI).
The second bound in~\eqref{eq:steepness_inequality} can be proved by applying the chain rule. Indeeed, the output chemical potential in the full inverter, Fig.~\ref{fig:inverter} a), can be envisaged as the function 
\begin{equation}
    \mu_O\left(\mu_I,\mu_E(\mu_I)\right),
\end{equation}
and we can use the chain rule to obtain the total derivative of  $\mu_{O}$ with respect to  $\mu_{I}$  
\begin{equation}
\begin{split}
   & \left|\frac{d\mu_{O}}{d\mu_{I}}\right|_{\text{full}} = \left|\frac{\partial \mu_O}{\partial \mu_I} + \frac{\partial \mu_O}{\partial \mu_E} \frac{d\mu_E}{d\mu_I}\right|\\
    \\
    &\le  \left|\frac{d\mu_{O}}{d\mu_{I}}\right|_{\text{simpl}} + \left| \frac{\partial \mu_O}{\partial \mu_E}\right|\left|  \frac{d\mu_E}{d\mu_I}\right|.
\end{split}
\label{eq: chain rule}
\end{equation}
$\left| \frac{\partial \mu_O}{\partial \mu_E}\right|$, in analogy with $\left|\frac{d\mu_{O}}{d\mu_{I}}\right|_{\text{simpl}}$, is $\le  \tanh\left(\frac{\Delta\mu}{4}\right) $.
Moreover, at steady state one must have
\begin{equation}
    \mu_{IE} = \mu_I + \mu_E \implies d\mu_{IE} = d\mu_I + d\mu_E 
\end{equation}
plus the relation coming from the conservation law
\begin{equation}
    dn_E + dn_{IE} = 0 \implies \langle n_E\rangle \,d\mu_E + \langle n_{IE}\rangle\,d\mu_{IE} = 0.
\end{equation}
The two together leads to
\begin{equation}
   \left| \frac{d\mu_E}{d\mu_I}\right| = \frac{1}{ 1+ \frac{\langle n_{E}\rangle  }{\langle n_{IE}\rangle}} \le 1.
\end{equation}
Putting everything back into Eq.~\eqref{eq: chain rule}, we conclude
\begin{equation}
    \left|\frac{d\mu_{O}}{d\mu_{I}}\right|_{\text{full}}\le 2.
\end{equation}
%Extrapolating from those two example, the intuition is that the inverter steepness is bound by the number of ways in which $I$ influences $O$ or  equivalently  by the effective power of  the input concentration which regulates the output concentration 

\section{Accuracy of the Gaussian approximation}
\label{app: accuracy rate function}
In this  Appendix, we show that the Gaussian approximation of the rate function is not sufficient to calculate first order moments up to order $1/V$, like $\langle x\rangle = \frac{\langle n_S\rangle}{V}$, while it is for second order moments, like $\text{cov}(x,y) = \frac{\text{cov}(n_S,n_D)}{V^2}$, up to the same accuracy. 
For simplicity, we limit the discussion to the 1D case. Analogous  results can be found for the 2D case, which is our case. \\
Consider a system described by an intensive variable $x$ that in our case would be the concentration of the species and by a scale parameter V, the volume.
Assume, as in our situation, that the \emph{exact} steady state distribution has the following scaling behavior
\begin{equation}
    P_{\rm ss}(x) = e^{-V I(x,V)}
\end{equation}
	 with 
  \begin{equation}
     I(x,V) = I_0(x) + \frac 1V I_1(x) + \frac{1}{V^2}I_2(x) +\dots.
     \label{eq:rate function expansion}
  \end{equation}
  The first term on the right-hand side, $I_0(x) = \lim_{V\to\infty} -\frac{1}{V}\log P_{\rm ss}(x) $, is called the rate function and can be derived from the master equation of the process by inserting $P_{\rm ss}(x)\asymp e^{-V I_0(x)}$ and ignoring terms of lower order in $1/V$.  The value $x^*$, where the rate function  has its minimum $I_0(x^*)$, represents the deterministic value approached in the limit $V\to \infty$.
  The Gaussian approximation consists in retaining only the rate function    $I_0(x)$ from the right-hand-side of Eq.~\eqref{eq:rate function expansion} and expanding it up to second order around $x^*$
 \begin{equation}\begin{split}
      P_{\rm ss}(x) \simeq P_{\text{Gauss}}(x)&\\ = \sqrt{\frac{VI_0''(x^*)}{2\pi}} &\exp\left(-\frac 12 V I_0''(x^*)\delta x^2\right) + O\left(\frac{1}{\sqrt{V}}\right)
 \end{split}
 \end{equation}
		with $\delta x = x -x^*$.
	We now show that this approximation is too drastic if one wants to  calculate $\langle \delta x\rangle$ with accuracy $1/V$. We do so by pointing out that the next order correction gives nonnegligible contributions. 	\\
   To obtain a probability distribution correct up to order $O\left(1/V^{3/2}\right)$,  the third order expansion of $I_0(x)$ and the term $\frac 1V I_1(x)$ are needed.  If we keep them, the improved approximation reads
\begin{equation}
    P_{\rm ss}(x) \simeq e^{-I_1(x) -\frac{1}{6}V I_0'''(x^*)\delta x^3} P_{\text{Gauss}}(x) +  O\left(\frac{1}{V^{3/2}}\right).
\end{equation}
In the relevant region where $ P_{\text{Gauss}}(x)$ is not zero, which is increasingly smaller as $V\to \infty$, we can approximate $I_1(x) \simeq I_1(x^*) + I_1'(x^*)\delta x$. Moreover, in the same region, we can bring down the argument of the exponential since it is $\ll 1$. From the normalization condition, it is $I_1(x^*)=0$ and we can write
\begin{equation}
\begin{split}
       & P_{\rm ss}(x) \simeq  \\ &\left[1 - I_1'(x^*)\delta x -\frac{1}{6}V I_0'''(x^*)\delta x^3\right]P_{\text{Gauss}}(x) +  O\left(\frac{1}{V^{3/2}}\right).
\end{split}
\end{equation}
If we use it to calculate $\langle\delta x\rangle$, we get
\begin{equation}
\begin{split}
    \langle \delta x\rangle & =  - I_1'(x^*) \langle \delta x^2\rangle_{\rm G}-\frac{1}{6}V I_0'''(x^*) \langle \delta x^4\rangle_{\rm G}\\
   & =-\frac{1}{V}\left(  \frac{I_1'(x^*)}{I_0''(x^*)} + \frac{1}{2}\frac{I_0'''(x^*)}{(I_0''(x^*))^2}\right)
\end{split}
\end{equation}
where the mean values $\langle \cdot \rangle_{\rm G}$ are computed with respect to $P_{\rm Gauss}(x)$.
Thus, we realize that $\frac 1V I_1(x)$ and the third order expansion of $I_0(x)$ give relevant contributions of order $1/V$. On the other hand, if we are interested in $\langle \delta x^2\rangle$, $P_{\text{Gauss}}(x)$ is sufficient as the next order correction gives zero contribution:
\begin{align}
        \langle \delta x^2\rangle &  =  \langle \delta x^2\rangle_{\rm G}- I_1'(x^*) \langle \delta x^3\rangle_{\rm G}-\frac{1}{6}V I_0'''(x^*) \langle \delta x^5\rangle_{\rm G}         \nonumber \\
       & = \langle\delta x^2\rangle_{\rm G}.
\end{align}
Notice that if thanks to the symmetry of the problem, $P_{\rm ss}(x)$ is known to be even with respect to $x^*$, then $I_1'(x^*) =I_0'''(x^*) = 0 $ and $P_{\text{Gauss}}(x)$ is accurate enough for calculating $\langle x\rangle$. This is what happens in the electronic Maxwell demon \cite{electronic} if all the transistors involved are assumed to have the same parameters.

\section{obtaining $J_{S_-\to S_+} $ in terms of $\text{cov}(n_S,n_D)$ }
\label{app: connection with corr}
We can derive a relationship between $J_{S_-\to S_+} $ and $\text{cov}(n_S,n_D)$ from Eq.~\eqref{eq:steady state condition 1}. To do so, we can substitute in it $\langle n_S n_D\rangle = \langle n_S\rangle \langle n_D\rangle + \text{cov}(n_S,n_D) $,  $\langle n_S \rangle = \langle \delta n_S\rangle +n_S^* $ and $\langle n_D \rangle = \langle \delta n_D\rangle +n_D^* $. Keeping in mind that, in the limit $V\to \infty$,  $\langle \delta n_S\rangle \ll n_S^* $, $ \langle \delta n_D\rangle \ll n_D^* $ and $\Delta \mu_S \ll 1$, we obtain to first order 
 \begin{equation}
   \langle \delta n_S\rangle = \frac{ k_+^S n_S^*\Delta\mu_S -  k_-^S\text{cov}(n_S,n_D)/V }{k_{\text{relax}}^S}.
    \label{eq: steady state 1 order 1}
\end{equation}
Inserting this value again in Eq.~\eqref{eq:steady state condition 1}, we have
\begin{equation}
\begin{split}
    J_{S_-\to S_+}= k^S_+\left(\langle n_S\rangle -Ve^{\mu_{S_+}}\right)\\
    \simeq k^S_+(\langle \delta n_S\rangle -n_S^*\Delta\mu_S)
\end{split}
\end{equation}
which yields Eq.~\eqref{eq:current system correlation} in the text.
\section{Evaluation of $\text{cov}(n_S,n_D)$ }
\label{app: evaluation of correlations}
We here derive the guassian approximation of the stationary probability distribution. From that, we obtain the covariance. We adopt exactly the same methodology used in~\cite{electronic}.\\
The master equation of the 2D jump-process is 
\begin{equation}
    d_t P(\boldsymbol n) = \sum_\rho \lambda_\rho(\boldsymbol n - \boldsymbol{\Delta}_\rho )P(\boldsymbol n - \boldsymbol{\Delta}_\rho ) - \lambda_\rho(\boldsymbol n)P(\boldsymbol n).
\end{equation}
 $\boldsymbol n  =(n_S,n_D)$, the index $\rho$ represents the possible transitions, and $\boldsymbol{\Delta}_\rho$ is the net change in the state of the Maxwell demon as a result of the transition $\rho$. There are two possible transitions that change the number of molecules in the system, $\boldsymbol{\Delta}^S_{\pm} = (\pm 1,0)$, and two possible transitions changing the number of molecules in the demon, $\boldsymbol{\Delta}^D_{\pm} = (0,\pm 1)$.\\
 As a preliminary step, we can substitute $P_{\rm ss}\asymp e^{-VI_0(\boldsymbol x)}$ in the stationary master equation, with $\boldsymbol x = \boldsymbol n/V$. One gets disregarding terms sublinear in the volume
\begin{equation}
    \sum_\rho \omega_\rho(\boldsymbol x) \left(  e^{\boldsymbol\Delta_\rho\cdot\nabla I_0(\boldsymbol x) } -1 \right) =0
    \label{eq: I in ME}
\end{equation}
where $\omega_\rho(\boldsymbol x) = \lambda_\rho(\boldsymbol n)/V$ are the rescaled rates. Then, we can proceed substituting the gaussian approximation  $I_0(\boldsymbol x) =\frac 12 \boldsymbol{\delta x}^T C^{-1}\boldsymbol{\delta x}$ in Eq.~\eqref{eq: I in ME}, with $\boldsymbol{\delta x} =\boldsymbol x - \boldsymbol x^*$. We obtain the following equation for the covariance matrix $C$
\begin{equation}
    CA + A^TC + B = 0
    \label{eq:matrix equation}
\end{equation}
with 
\begin{equation}
    A_{ij} = \sum_\rho \partial_{x_i}\omega_\rho(\boldsymbol x^*) (\boldsymbol\Delta_\rho)_j = \left(\begin{matrix}
        -k_{\text{relax}}^S & -\alpha k_{\text{relax}}^D \\
         0 & -k_{\text{relax}}^D,\\
    \end{matrix}\right)
\end{equation}
and 
\begin{equation}
    B_{ij} = \sum_\rho \omega_\rho(\boldsymbol x^*)(\boldsymbol\Delta_\rho)_i(\boldsymbol\Delta_\rho)_j = \left(\begin{matrix}
        2\frac{n_S^*}{V}k_{\text{relax}}^S & 0 \\
         0 & 2\frac{n_D^*}{V}k_{\text{relax}}^D\\
    \end{matrix}\right).
\end{equation}
Eq.~\eqref{eq:matrix equation} can be explicitly solved for $C$ yielding
\begin{equation}
    C_{12} = \frac{\text{cov}(n_S,n_D)}{V^2} =  -\frac{\alpha n_S^*}{V^2}\left(\frac{k_{\text{relax}}^D/k_{\text{relax}}^S}{k_{\text{relax}}^D/k_{\text{relax}}^S+1 }\right).
\end{equation}
\section{Information flow}
\subsection{$\dot{\mathcal{I}}$ in the limit of a fast demon}
\label{app:evaluation of I}
We obtain Eq.~\eqref{eq:info flow fast demon} starting from Eq.~\eqref{eq:info flow expressions}. We perform calculations setting $\Delta\mu_S =0$ as justified in the text.
\begin{align}
     \dot{\mathcal{I}} &=-\dot{\mathcal{I}}_S \label{eq: app info flow expression}
\\
     &=- \sum_{n_D,n_S} j_{ss}(n_S+1\leftarrow n_S|n_D)\log \frac{P_{\rm ss}(n_D|n_S+1)}{P_{\rm ss}(n_D|n_S)}.\nonumber
\end{align}
In this regime,  $P_{\rm ss}(n_D|n_S)$ is a Poissonian with mean  $ n_D(n_S)$ given by Eq.~\eqref{eq:demon_input_output}. In the large volume limit, $ n_D(n_S)$ can be approximated by the linear relation Eq.~\eqref{eq:linear relation}. This leads to 
\begin{equation}
\begin{split}
    &\log \frac{P_{\rm ss}(n_D|n_S+1)}{P_{\rm ss}(n_D|n_S)} = \log \left[e^\alpha \left(1-\frac{\alpha}{n_D(n_S)}\right)^{n_D}\right] \\ 
    &\simeq \alpha \frac{n_D(n_S) -n_D  }{n_D(n_S)} -\frac 12\frac{\alpha^2 n_D}{n_D(n_S)^2}\\
    &\simeq -\alpha \frac{(n_D -n_D^*) + \alpha(n_S-n_S^*)}{n_D^*} -\frac{1}{2}\frac{\alpha^2}{n_D^*}.
\end{split}
\end{equation}
The last term on the right-hand side is a constant and it gives zero contribution to Eq.~\eqref{eq: app info flow expression} since 
\begin{equation}
    \sum_{n_D,n_S} j_{ss}(n_S+1\leftarrow n_S|n_D) = 0.
\end{equation}
This equality comes from the fact that the barycenter of a steady state distribution does not move horizontally.
The  term $-\alpha \frac{(n_D -n_D^*) + \alpha(n_S-n_S^*)}{n_D^*}$ enters in Eq.~\eqref{eq: app info flow expression} together with 
\begin{align}
         j_{ss}(n_S+1\leftarrow n_S|n_D) &\simeq \\  k^S_{\text{relax}}[n_S^*P_{\rm ss} &(n_S ,n_D)- n_SP_{\rm ss}(n_S+1,n_D)].\nonumber
\end{align}
%we remind $k^S_{\text{relax}} =\left(k_+^S +  k_-^S\frac{n_D^*}{V}\right)$. 
To obtain the final result 
\begin{equation}
    \dot{\mathcal{I}} = \alpha^2 \frac{n_S^*}{n_D^*} \,k^S_{\text{relax}},
\end{equation}
one  takes the average with respect to $n_D$, firstly, and with respect to $n_S$, secondly: $\langle n_D\rangle$ is given by the linear relation~\eqref{eq:linear relation} and $\langle n_S\rangle$ can be approximated in this context with $\langle n_S\rangle \simeq n_S^*$.
\subsection{Scaling of $\dot{\mathcal{I}}$ in the limit of a slow demon}
\label{app: scaling I}
We call $\epsilon = k_{\text{relax}}^D/ k_{\text{relax}}^S$. With $\Delta\mu_S =0$, one has 
\begin{equation}
   \lim_{\epsilon\to 0} P_{\rm ss}(n_S|n_D) =  f(n_S).
\end{equation}
Put into words: the rates of the system are infinitely faster than those of the demon, thus, for any given $n_D$, the system equilibrates with respect to $n_S$ to a distribution that, due to $\Delta\mu_S = 0$, is actually independent of $n_D$. In particular,
$f(n_S) $ is a Poissonian with average $\langle n_s\rangle = e^{\mu_{S_-}} =e^{\mu_{S_+}}$.
 Therefore, in this limit, the steady state probability ditribution factorizes and we can write to next order in $\epsilon$
\begin{equation}
    P_{\rm ss}(n_S,n_D) = f(n_S)g(n_D) + \epsilon \,h(n_S,n_D).
    \label{eq: prob distr slow demon}
\end{equation}
The information flow can be evaluated from Eq.~\eqref{eq:info flow expressions} as
\begin{equation}
      \dot{\mathcal{I}} = \dot{\mathcal{I}}_D  = \sum_{n_S} \sum_{n_D'>n_D} j(n_D'\leftarrow n_D|n_S)\log \frac{P_{\rm ss}(n_S|n_D')}{P_{\rm ss}(n_S|n_D)}.
\end{equation}
Around the working point, from where the major contributions to $\dot{\mathcal{I}}$ come from, the current along $D$ scales
\begin{equation}
    j_{ss}(n_D'\leftarrow n_D|n_S) \propto k_{\text{relax}}^D
    \label{eq: scaling jD}
\end{equation}
and from Eq.~\eqref{eq: prob distr slow demon},
\begin{equation}
     P_{\rm ss}(n_S|n_D) = f(n_S) + \epsilon \,\frac{h(n_S,n_D)}{g(n_D)}
\end{equation}
so that 
\begin{equation}
    \log \frac{P_{\rm ss}(n_S|n_D')}{P_{\rm ss}(n_S|n_D)} \propto \epsilon
    \label{eq: scaling log P}.
\end{equation}
Combining Eq.~\eqref{eq: scaling jD} and \eqref{eq: scaling log P}, we have 
\begin{equation}
    \dot{\mathcal{I}} \propto  k_{\text{relax}}^D\left(\frac{k_{\text{relax}}^D}{k_{\text{relax}}^S}\right)
\end{equation}
as claimed in the text.

\bibliography{biblio}% Produces the bibliography via BibTeX.

\end{document}